\documentclass[12pt]{article}
\usepackage{graphicx,rotating}
\usepackage{bm} % bold math
\usepackage{amssymb,amsmath}

\newcommand{\fsi}{FSI }
\newcommand{\degr}{$^o\,$}
\newcommand{\rts}{ \sqrt s}

\begin{document}

\thispagestyle{empty}

\title{$\eta$ bound state: theory and experiment}

%\author{\firstname{Hartmut}\lastname{Machner}
%\inst{1}\fnsep\thanks{\email{hartmut.machner@uni-due.de.de}}}

%\institute{
%Fakult\"{a}t f\"{u}r Physik, Universit\"{a}t Duisburg-Essen, Lotharstr. 1, %47048 Duisburg, Germany }

\author{\textbf{H. Machner\footnote{hartmut.machner@uni-due.de}}\\Fachbereich Physik, Univ. Duisburg-Essen,\\ Lotharstr. 1, 47048 Duisburg, Germany}

\maketitle

\begin{abstract}
%%\abstract{%
The search for a quasi bound $\eta$ meson in atomic nuclei is reviewed. This tentative state is studied theoretically as well as experimentally.\end{abstract}
%%}
%%\maketitle

%%\pacs{21.85.+d, 13.75.-n}

%

\section{Introduction}\label{sec:Introduction}

This report is based on a recent review by the author \cite{Machner14}. Other reviews covering the topic in part are in \cite{Kelkar13} and \cite{Krusche14}. Atomic nuclei are built up by nucleons. The quark substructure is not visible. The nucleons are bound together by the strong force. The replacement of nucleons by $\Lambda$- or $\Sigma$-hyperons was successfully studied, by producing them via recoil-free kinematics, i. e. they are produced at rest. Bound system consisting of an atomic nucleus and another negatively charged particle, where binding appears due to the Coulomb interaction, are atoms.  The successful method of producing pionic atoms was again recoil free kinematics.  We will come back to this point.

Since the $\eta$ is electrical neutral, binding can occur only via the strong interaction.  The observation of an $\eta$ bound state would be the first time that a boson is bound in a nucleus.  Since a boson is not effected by the Pauli principle, it can be in a state where the nucleon density is maximal. Observation of such a state would allow to study the in medium properties of bound objects. The $\eta$ is short lived ($\tau =5*10^{-19}$ s) and therefore secondary beams of the are impossible.

It is well known that the Delta resonance $P_{33}(1232)$ plays a dominant role in case of pion production.  It seems that the $S_{11}(1535)$ plays a similar role in case of $\eta$ production although cross sections in this case are much smaller.

The scattering lengths of
the pion-nucleon interaction is rather small and as a result the
strong interaction shift in the $1s$ state of pionic atoms is
repulsive. Contrary to this, the $\eta$-nucleon interaction at small
momenta is attractive and rather strong. This was first pointed out
by Bhalerao and Liu \cite{Bhalerao85} and later applied by Haider
and Liu \cite{Haider_Liu86} to predict quasi bound $\eta$ mesons in
atomic nuclei for mass numbers $A\geq 12$. In the following text we
apply the standard sign convention in meson physics
\cite{Goldberger-Watson} for the $s$ wave scattering parameters
$p \cot \delta_0 = 1/a + 1/8( r_0p^2)\label{definition}
$ with $p$ the $\eta$ momentum, $\delta_o$ the $s$ wave phase shift, $a$ the scattering length and $r_0$ the effective range. For a real attractive
potential $a_{r} < 0$ means binding. Contrary to the $\pi-N$ systems
where the scattering length is real at very small energies here the
$\eta N\to \pi N$ channel is always open and hence the scattering
length is complex. From such large  values for the scattering length
$a(\eta N)$, Haider and Liu\cite{Haider_Liu86} have shown that
$\eta$ can be bound in nuclei with  A $\ge$ 12. In the following text we frequently use the
term bound state instead of the more strict quasi bound state. This is common in the literature.

\section{Theoretical Considerations}\label{sec:Theoret-Considerations}
A state is called a bound state in the usual sense when the sum of its constituent masses is larger than the mass of the composite. In non-relativistic quantum mechanics binding is represented by an attractive potential and the state is a solution of the radial Klein-Gordon equation. These solutions lie on the imaginary axis in the momentum plane with $\text{Im}(p)=p_i>0$
However, a possible $\eta$ bound state is not stable since always the interaction
\begin{equation}\label{eq:eta-decay}
\eta+N\to \pi+N'
\end{equation}
with a nucleon $N$ is possible. If the $\eta$ bound state was in a $s$-state the energy of the final state is $$m_\eta+m_N-B_\eta=m_\pi+m_{N'}+T_\pi+T_{N'}$$ with $T$ the kinetic energies in the final state and $B_\eta$ the binding energy. Here we have neglected Fermi motion of the nucleon and the recoil of the residual nucleus. Assuming a binding energy $B_\eta=10$~MeV, this leads to $T_\pi\approx 317$~MeV and $T_{N'}\approx 47.3$~MeV. These energies are clearly too large for the two final state particles to stay in the nucleus. Because of the possible decay of the state it is a quasi bound state and this fact is accounted for by a complex potential. The task is now to produce a complex potential for elastic scattering $\eta N\to \eta N$, construct from this a complex $\eta A\to \eta A$ potential and then search for poles in the upper part of the second quadrant in the complex plane.

The $\eta$-nucleon scattering length $a(\eta N) $ or more generally
the matrix $T(\eta N\to \eta N)$  is quite poorly known. As stated above the lifetime of $\eta$'s is short, so $a(\eta N)$ or $T(\eta N\to \eta N)$ has to be extracted
in rather indirect ways. The inputs are production cross sections of
$\pi^-p\to \eta n$ and $\gamma p\to \eta p$ reactions. Also decays
into the channels $\gamma N$, $\pi N$, $\pi\pi N$ and $\eta N$ were
considered. The major mechanism that generates the imaginary part of
$a(\eta A)$ is the reaction $\eta A_i\to N^*(A - 1) \to \pi A_f$,
where $N^*$ is the nucleon resonance $N^*(1535)$ with a strong
coupling to both the $\eta$ and the pion \cite{PDG10},
\cite{Wycech95}. The deduced values for the scattering length range from $0.22 +  i0.235$ fm to $1.14+i0.31$ fm.

The standard approach is to construct from the $\eta$-nucleon scattering length an optical potential for the $\eta$-nucleus interaction with $A$ the mass number of the nucleus, and then to solve a wave equation with this potential \cite{Haider_Liu86}, \cite{Friedman13}, \cite{Wycech95}, \cite{Wilkin93}. The complex optical potential is given by
\begin{equation}\label{equ:Optical}
U_\text{opt} = V+iW=-\frac{2\pi}{\mu} T(\eta N\to \eta N) A \rho(r)
\end{equation}
with $\mu$ the reduced $\eta N$ mass, $T(\eta N\to \eta N)$ the $\eta$-nucleon transition matrix and $\rho$ the nuclear density. We will call this relation as the $T\rho$ approximation. In the impulse approximation the relation
\begin{equation}\label{equ:Tscatterin}
a(\eta N) = T(\eta N\to \eta N,\rts_0)
\end{equation}
with $$\sqrt{s_0} =m_\eta+m_N$$ holds. For a bound state one needs to  know the $T$ matrix at $$\sqrt{s} = \sqrt{s_0}-B_\eta $$ i.e. below threshold.

Different groups employed different wave equations to search for poles. This leads of course to different results. Another source of ambiguities are the different $\eta$ nucleon scattering lengths.

Here we study the importance of the input on the final result. We compare the mass dependence of the binding energy and the width  for the two extreme values of the scattering length. Such a comparison was made in Ref. \cite{Friedman13}.
\begin{figure}[h]
\includegraphics[width=0.45\textwidth]{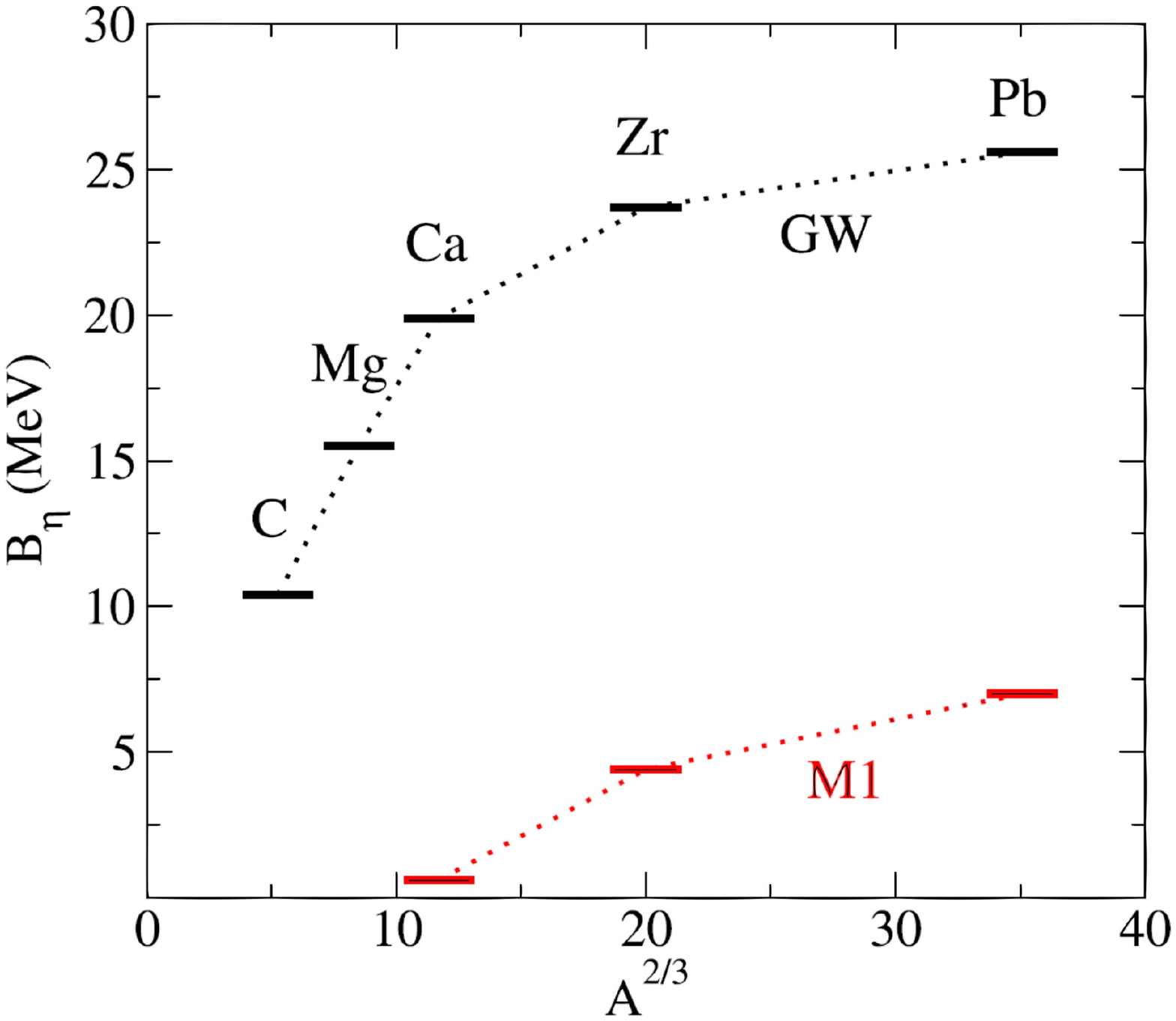}
\includegraphics[width=0.45\textwidth]{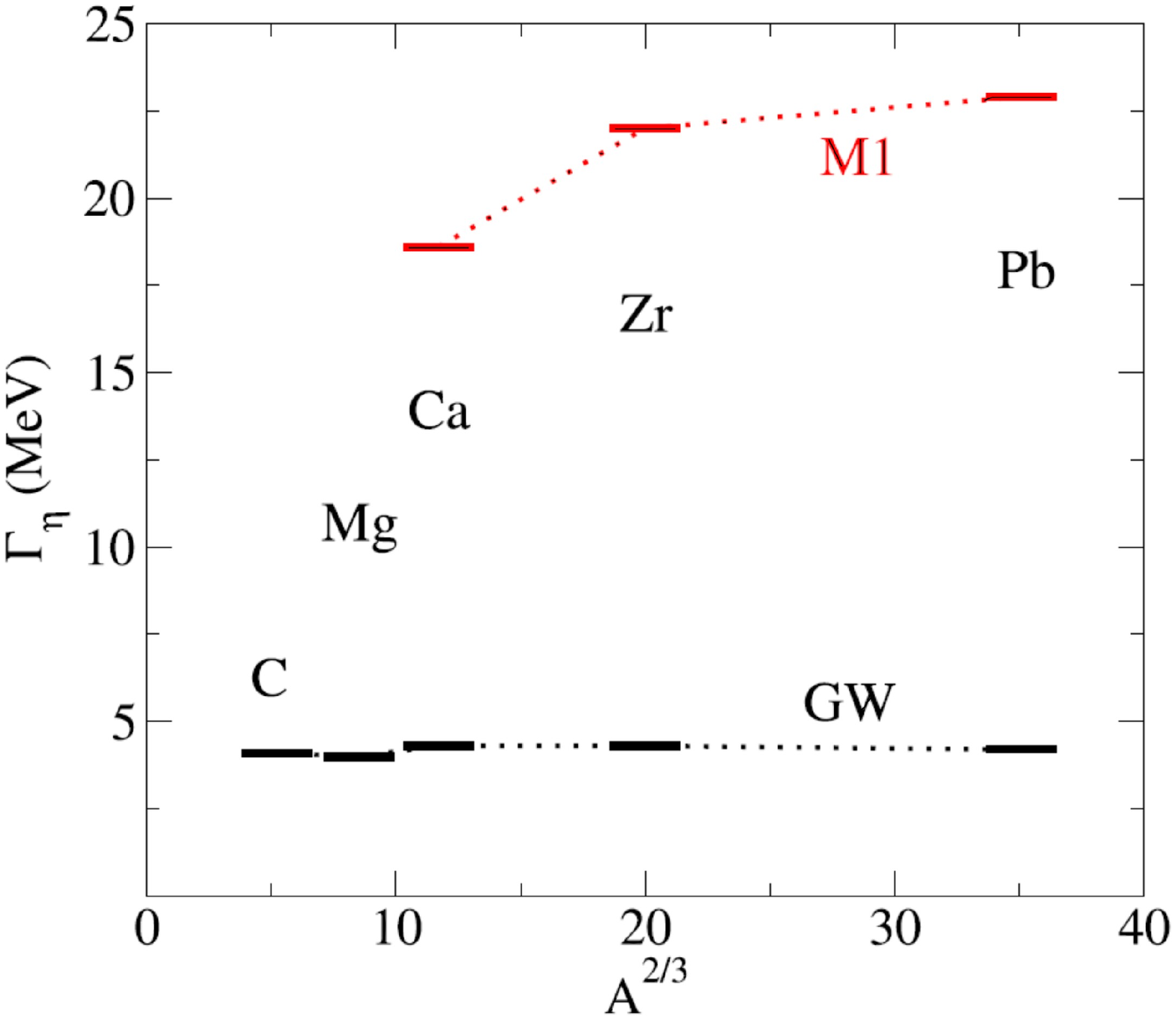}
\caption{Theoretical calculations of the complex energies for $\eta$-bound nuclei in $1s$ state(from \cite{Friedman13}). The results labeled with GW use the Green-Wycech \cite{Green05} result of the scattering length $a(\eta N)=0.97+i0.26$ fm,  while those labeled M1 use the result from Mai et al. \cite{Mai12} $a(\eta N)=0.22+i0.24$ fm  Left frame: the binding energy, right frame the width.}
\label{Fig:Gal}
\end{figure}
The results doe the smallest value of the scattering length is from \cite{Mai12} and the largest from Ref. \cite{Green05}. The results are shown in Fig. \ref{Fig:Gal}. The larger real part of the scattering length leads to rather strong binding. However, the imaginary parts although are almost identical lead to strongly different widthes. The Green-Wycech results gives an almost mass independent width.

A possible way to extract the properties of a bound state is to extract the $\eta$ nucleus scattering length from the final state interaction \cite{Goldberger-Watson}. One has to measure an excitation function of a
reaction
\begin{equation} ^{Z_1}A_1+^{Z_2}A_2\to
\{^{Z_1}A_1+^{Z_2}A_2\}_{gs}+\eta
\end{equation}
with $\{^{Z_1}A_1+^{Z_2}A_2\}_{gs}$ the fused nuclear system in its ground state and the $\eta$ relative to that in a $s$ state. One can either measure the $\eta$ or the fused nuclear system. The measurement of the decay of the $\eta$ into photons does not allow the conclusion that the nuclear system is in its ground state due to the limited resolution in the two photon detection. Instead one measures the four vector of the fused nucleus together with particle identification and reconstructs the properties of the $\eta$. This, however, limits the method to light nuclei.

The method is to extract the effective range parameters from the matrix element
\begin{gather}\label{equ:cross_section}
|f_s|^2=\frac{d\sigma_s}{d\Omega}\frac{p_i}{p_f}
\end{gather}
with $p$ the momenta in the incident and final state in the cm system and $d\sigma_s/d\Omega$ the $s$ wave part of the cross section, as will be discussed in the next section.  The parameters scattering length $a$ and effective range $r_0$ have to be complex since always the channel $\eta+N\to\pi+N$ is open. Because the square of $a$ is fitted to the data the sign of $a_r$ can not be found from such measurements. The case with more than one $s$ wave will be discussed below.

One can naively assume that the $s$ wave part of the cross section close to threshold is just $d\sigma_s/d\Omega = \sigma_{tot}/4\pi$.
However, often other waves than just the $s$ wave contribute to the total cross section even close to threshold. In this case the decomposition of the total cross section into partial waves is possible from the knowledge of  spin observables in addition to cross sections.
In the following paragraphs we will give some theoretical prerequisites allowing to extract the $s$ wave contribution from measurements.

In the simplest approach the $s$ wave amplitude is related to the
scattering length via
$f_s(p)=f_B/(1-iap)
\label{equ:scattering_length}$
with $f_B$ the production amplitude.
$f_s(p)$ has a pole in the complex plane that occurs for
$$p_0=\frac{-i}{a}=\frac{-ia^*}{|a|^2} \label{equ:pole_position}\,.$$
With $$E=\frac{p_{0}^2}{2\mu_{\eta A}}$$ we find
$$B_\eta =  \frac{a_r^2-a_i^2}{2\mu_{\eta A}|a|^4)} \label{equ:B-pole}$$
and
$$\frac{-\Gamma_\eta}{2} =  \frac{2a_ra_i}{2\mu_{\eta A}|a|^4} \label{equ:Gamma-pole}\,.$$
For $B_\eta >0$ follows
$|a_r|>|a_i|$ and $a_r a_i<0$. Unitarity requires $a_i>0$ and therefore $a_r<0$.

\section{Experiments}\label{sec:Experiment}
The experimental searches for bound or quasi bound states is not a story of successes. Early experiments \cite{Chrien88} were later shown not be be quasi free \cite{Nagahiro09} and hence no effect could be seen. Other experiments \cite{Lieb88}, \cite{Sokol99} and \cite{Baskov12} were not conclusive, See Ref. \cite{Machner14} for a detailed discussion.

A photoproduction experiment was performed at the MAMI accelerator in Mainz making use of the TAPS spectrometer \cite{Pfeiffer04}. A tagged photon beam with 800 MeV maximum energy on a $^3$He target was used.  The reaction studied  was
\begin{equation}\label{equ:Taps_Reaction}
\gamma+{^3\text{He}}\to \pi^0+p+X
\end{equation}
as a function of $W$, which is the CM energy reduced by the deuteron mass and the {$^3$He} binding energy. An enhancement was found in the difference spectrum between the angular range 180\degr to 170\degr and 170\degr to 150\degr. The authors \cite{Pfeiffer04} claimed to have seen a bound $\eta$ state which implies that the first step $\gamma+^3\text{He}\to \eta\otimes ^3\text{He}$
occurred followed by
$\eta+p\to N^{+*}\to \pi^++p$.

\begin{figure}[h]
\begin{center}
\includegraphics[width=0.50\textwidth]{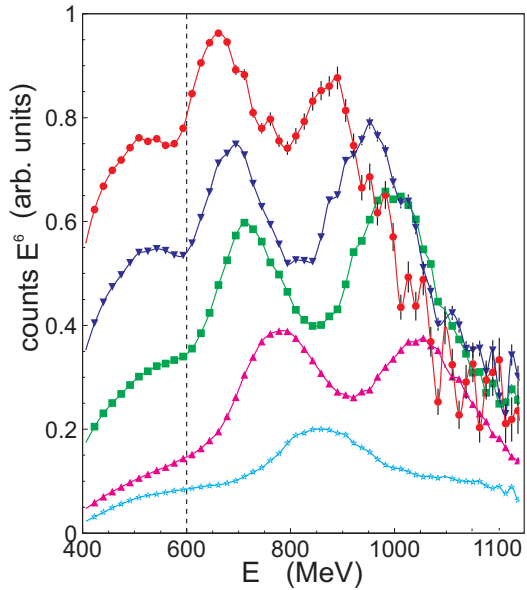}
\caption{Excitation functions of $\pi^0 p$ back-to-back pairs for different ranges of the opening angle $\theta_{\pi^0}+\theta_p$ after removal of the overall energy dependence $\propto E_\gamma^{-6}$. From top to bottom opening angle ranges of: 165\degr - 180\degr, 150\degr - 165\degr, 140\degr - 150\degr, 130\degr - 140\degr, and 120\degr - 130\degr. The vertical line indicates the $\eta$-production threshold. }
\label{Fig:Pheron}
\end{center}
\end{figure}
Almost the same group repeated the experiment with again the TAPS spectrometer plus the Crystal Ball detector \cite{Pheron12}. The experiment benefitted not only from the now almost 4$\pi$ acceptance but also from much higher statistics. The photon energies ranged from 0.45 GeV to 1.4 GeV. The result of this measurement is shown in Fig. \ref{Fig:Pheron}.
The strong rise of the $\pi^0 p$ cross section above the $\eta$ production threshold is similar to the previous experiment and supports the possibility of a resonance in the threshold region. However, the structures visible at higher energies have not been seen in \cite{Pfeiffer04}. They are in the so called second and third resonance region and their walk with angle is purely kinematical.

The WASA collaboration \cite{Adlarson13} studied the reaction $d+d\to \pi^-+p+{^3\text{He}}$ reaction. The idea is that an intermediate $\eta\,\alpha$ bound state might exist. The whole reaction chain is then
\begin{equation}\label{equ:WASA-dd}
d+d\to \eta\otimes\alpha\to N^*(1535)+{^3\text{He}}\to (\pi^- + p)+{^3\text{He}}\,.
\end{equation}
The deuteron beam momentum varied between 2.185 GeV/c and \- 2.400 GeV/c. No anomaly in the excitation function for beam momenta below and above threshold has been seen.

Experiments employing transfer reactions are favourable; in such experiments the whole beam momentum can be transferred to a nucleon or a cluster of nucleons. The remaining system then does not carry linear momentum and thus favours the probability that a produced $\eta$ is bound to the residual nucleus. This method, originally developed in the production of hypernuclei  \cite{Bruckner76}, was successfully applied in the study of pionic atoms \cite{Yamazaki96}. In order to transfer the beam momentum almost completely to the emerging particle it has to be emitted in the forward direction close at zero degree. One such experiment
\cite{Hayano97}, \cite{Gillitzer06} employed the GSI fragment separator. The search was done with the $d,^3he$ reaction. The spectrometer is flooded by break up protons having beam velocity and therefore the same magnetic rigidity $p/Z$ as the $^3$He particles of interest and are thus undistinguishable. So far no final result is published \cite{Gillitzer14}.

One nucleon transfer guarantees a rather large cross section. This is not the case for two nucleon transfer reaction $A+p\to (A-1)\bigotimes\eta+{^3He}$.  However, it is just this experiment by the GEM collaboration \cite{Budzanowski09} which claims to have observed an $\eta$ mesic bound state with sufficient significance. We will therefore discuss this experiment in more detail. The experiment made use of two signatures simultaneously: transfer reaction with recoil free kinematics and back to back emission of a pion and a nucleon from a possible reaction chain $\eta +N\to N*\to N'+\pi$ with the $N^*$ almost at rest.  A proton beam from the COSY J\"{u}lich accelerator with momentum of 1745 MeV/c was used, where $\eta$ mesic states with binding energies $-30\text{ MeV}\leq B_\eta\leq 0\text{ MeV}$ can be produced with a momentum transfer $q\leq$ 30 MeV/c.
\medskip
\begin{figure}[ht]
\begin{center}
\includegraphics[width=0.60\textwidth]{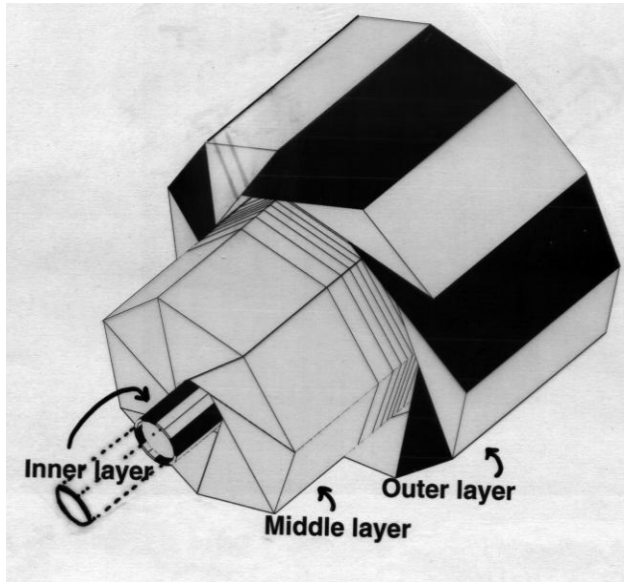}
\caption{The ENSTAR detector \cite{Betigeri07} surrounding the target. It consists of wedges from scintillating material. Read out is performed by scintillating fibres collecting the light in grooves milled in the wedges and transporting it to photo tubes.  One half of this detector is shown. The inner two layers are extruded for clarity.}
\label{Fig:ENSTAR}
\end{center}
\end{figure}
The high resolution magnetic spectrograph Big Karl \cite{Drochner98} was used to identify $^3$He ions and their momenta. The decay into a proton and $\pi^-$ with the two final particles emitted almost back to back to each other was measured with a dedicated detector ENSTAR \cite{Betigeri07}. It surrounds the target and one half of it is shown in Fig. \ref{Fig:ENSTAR}. By construction it is capable of determining azimuth and polar angle.
\begin{figure}[ht]
\begin{center}
\includegraphics[width=0.5\textwidth]{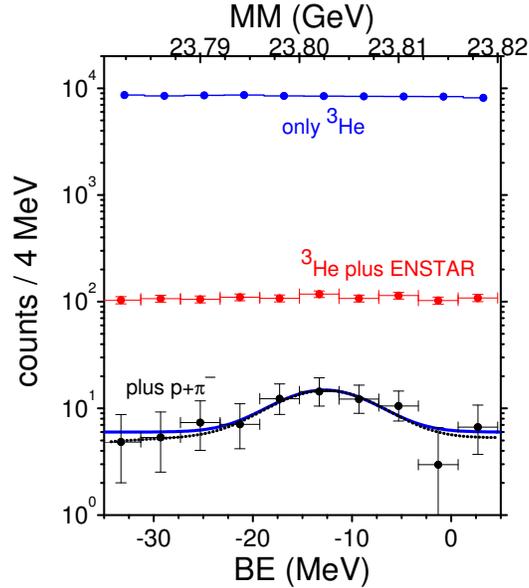}
\caption{Binding energy spectra. Upper curve: data with the requirement of a $^3$He in the focal plane. Middle curve: a coincidence between $^3$He and the ENSTAR detector. Lower curve: coincidence between $^3$He and a $\pi^-$ and proton being back to back (bb) emitted recorded in the ENSTAR detector. The solid curve is a fitted Gaussian together with a constant, the dashed curve a gaussian plus a  polynomial.}
\label{Fig:BB_Bg}
\end{center}
\end{figure}

We want now to discuss the effect of the conditions applied to missing mass spectrum or binding energy spectrum.  The momenta measured in the  FP are shown in Fig. \ref{Fig:BB_Bg}, converted to binding energy.
\begin{figure}[h!]
\begin{center}
\includegraphics[width=0.50\textwidth]{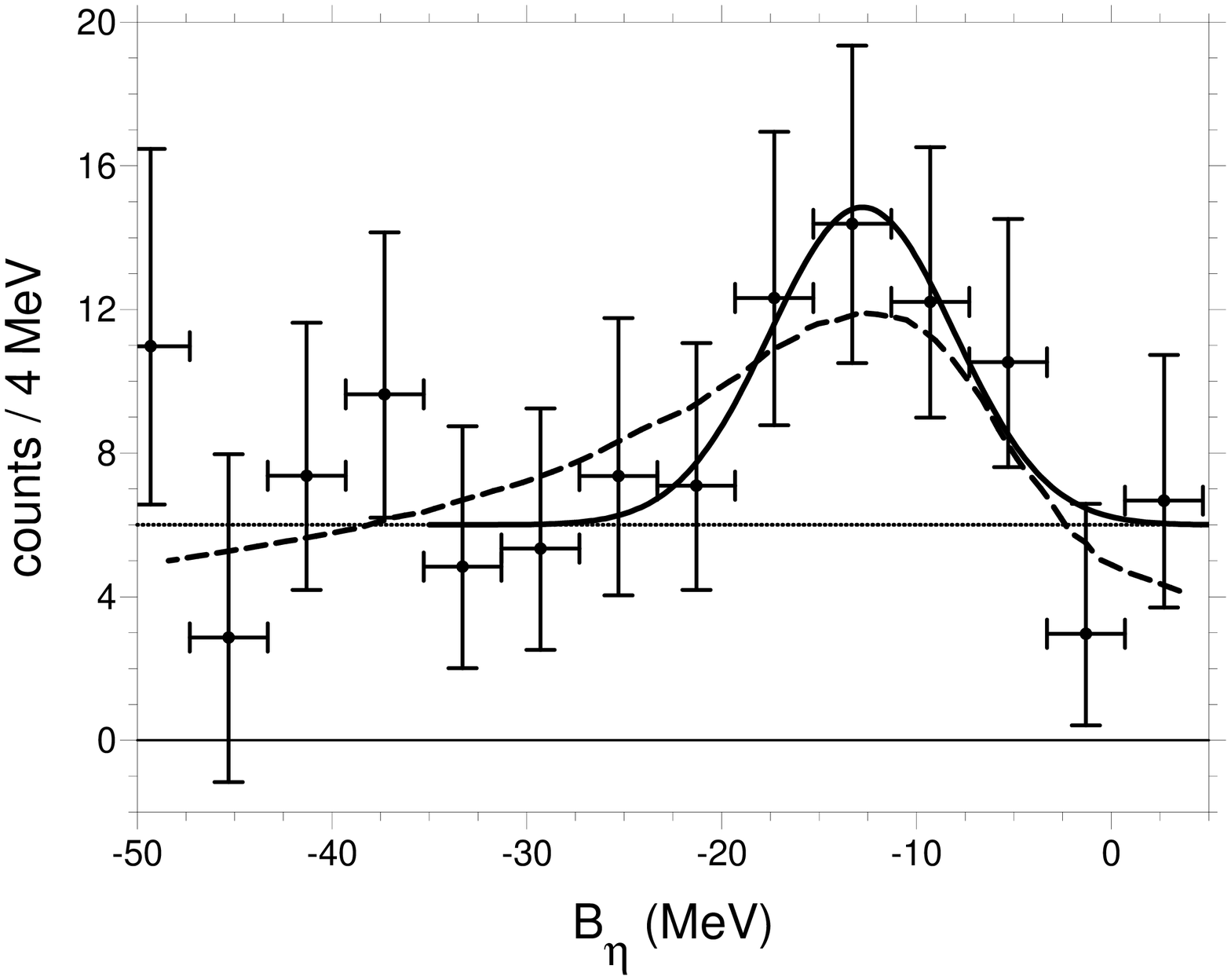}
\caption{The final binding energy spectrum. Note the expanded view and the different binning compared to Fig. \ref{Fig:BB_Bg}.  The data are shown with Poisson error bars. A fit to the data with a Gaussian and a constant background are shown (solid curve). A fit with a Breit-Wigner form with a coherent non resonant fraction is shown as dashed curve (from \cite{Haider-Liu10}).}
\label{Fig:Final}
\end{center}
\end{figure}
This spectrum shows a peak on a continuum. This continuum was parameterised by a constant as well as polynomials while for the peak a Gaussian was assumed. In addition fits were performed applying Poisson statistics. The significance of the peak is around 5$\sigma$ \cite{Budzanowski09}. The centroid $E_B$ and Gaussian width $\sigma$ were found to be -12.0$\pm$2.2 MeV and 4.7$\pm$1.7 MeV.

The procedure applied by the GEM collaboration \cite{Budzanowski09} to assume a further background below the peak was questioned by Haider and Liu \cite{Haider-Liu10}. The final state can also be reached by a non resonant reaction for which they used a microscopic-theory based nearly energy-independent amplitude. The need of adding non-resonant amplitude is further discussed in \cite{Liu14}. Then there will be an interference between this amplitude and the one for the resonant production. They fitted the corresponding amplitudes to the experimental data and found indeed a serious interference effect which shifts the calculated Breit-Wigner maximum towards the experimental maximum. The same is true for the width. One such fit is also shown in Fig. \ref{Fig:Final}. For this curve the $\eta N$ scattering length is (0.250 + 0.123i) fm.

Another method proposed to search for $\eta$ bound states is to
study the final state interaction (\fsi) between the $\eta$ meson
and a nucleus.

\begin{figure}[ht]
\begin{center}
\includegraphics[width=0.5\textwidth]{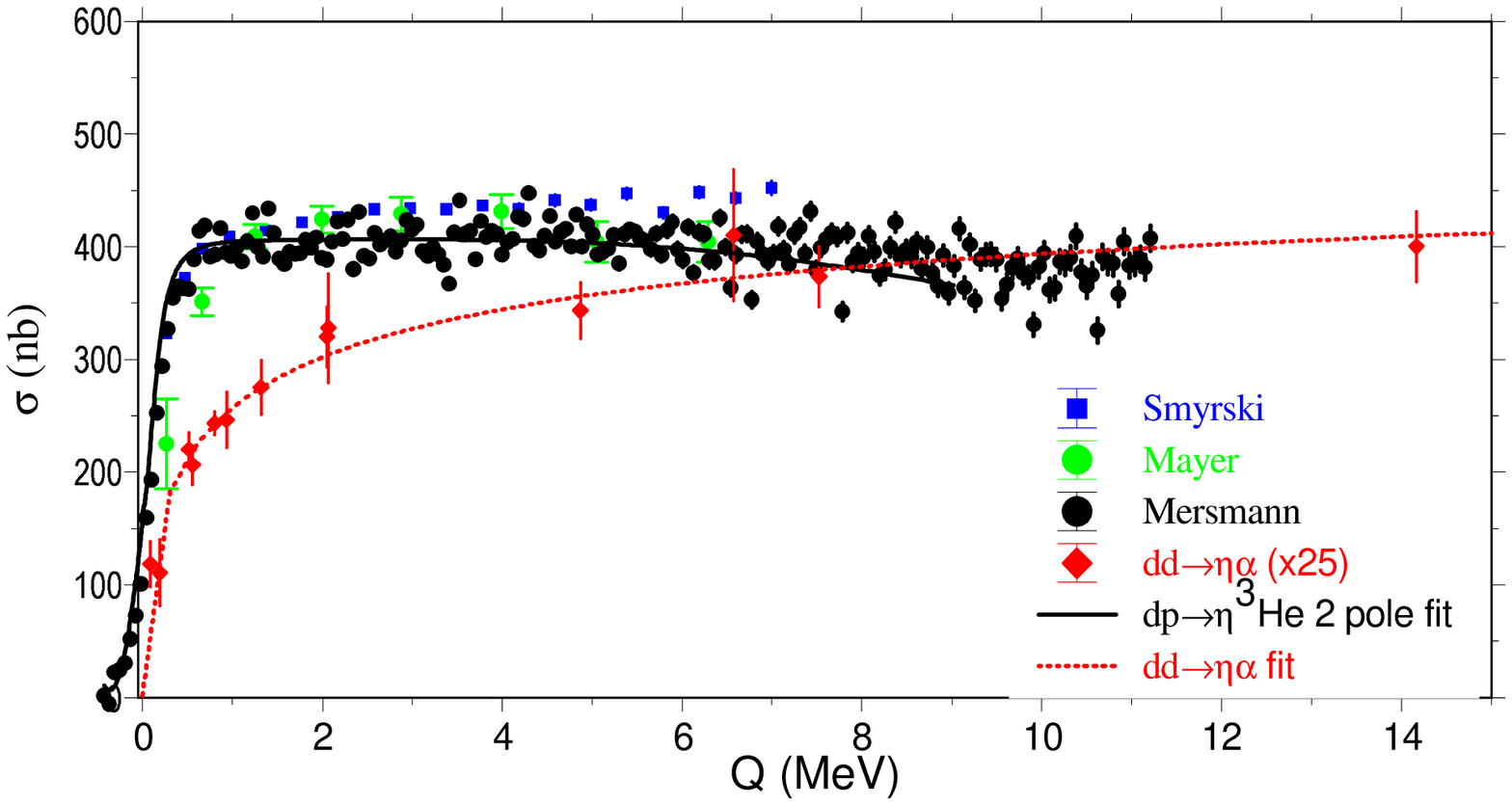}
\caption{Comparison of the excitation functions for the two reactions $d+p\to \eta+^3$He and $d+d\to \eta+^4$He. The total cross sections for the former reaction are from \cite{Mayer96} (diamonds), \cite{Smyrski07}(squares), and\cite{Mersmann07} (dots, five points together). The solid curve is the two pole fit. Data for the latter reaction (triangles up) are from Frascaria et al. \cite{Frascaria94}, Willis et al. \cite{Willis97}, Wronska et al. \cite{Wronska05}, and the GEM collaboration Budzanowski et al. \cite{Budzanowski09b}  The dotted curve is the scattering length fit to these data.}
\label{Fig:Compare_he3-He4}
\end{center}
\end{figure}

The system  most intensively studied is the $d+p\to ^3$He$+\eta$ reaction. Data are from Refs.  \cite{Berger88},  \cite{Betigeri00}, \cite{Adam07}, \cite{Rausmann09}, \cite{Mayer96}, \cite{Smyrski07}, \cite{Mersmann07}. Those close to threshold are shown in Fig. \ref{Fig:Compare_he3-He4}.
In this figure the recently published data are compiled. Obviously there are differences between the different data sets. This leads of course to different results for the final state parameters.
COSY~11 \cite{Smyrski07}, quoted also as Smyrski, and ANKE \cite{Mersmann07}, quoted also as Mersmann,  measured at COSY applying the internal deuteron beam. The momentum of the beam increased linearly with time. Data were taken continuously and later put into bins with widths $\Delta Q$. Details of corrections for nonlinearity within bins and correction due to finite beam resolution are given in \cite{Smyrski07} and \cite{Mersmann07a}. COSY~11 \cite{Smyrski07} applied Eq. (\ref{equ:scattering_length}) to their data and obtained
\begin{equation}
a_{\eta {^3He} } = {\pm}(2.9 {\pm} 2.7)+i(3.2 {\pm} 1.8) \text{ fm}.
\end{equation}
This corresponds to a possible bound state at $B_\eta=-0.2\pm 0.8$~MeV. So this result points more to a virtual than to a bound state. The half width is $\Gamma/2=1.9\pm 0.4$~MeV. However, when we repeated the fit for the data from Ref. \cite{Smyrski07} we found different values and moreover they depend on the fit interval. While the value for $a_i$ is quite stable, $a_r$ varied from $0.0\pm
6000$ fm, when the full data set is included in the fit, to $2.1\pm
2.7$ fm, when the range is limited to 2.2~MeV. The large error is an
indication that the option of fitting the full range is useless, because the assumption of pure $s$ wave is wrong. The
imaginary part is $3.6 \pm 1.2$ fm. These numbers are in agreement
with the published values. This finding is an indication
that already for excess energies above 2.2~MeV Eq.
(\ref{equ:scattering_length}) is no more applicable and the
effective range has to be considered in addition to the scattering
length as has been stressed in Ref. \cite{Niskanen13}. On the other hand the ANKE data show after the rapid rise a gentle
decrease with increasing energy. Mersmann \cite{Mersmann07a} has
performed  a corresponding fit to the ANKE data including the
smearing as discussed above. This fit yielded
$$
a_{\eta ^3He }  = \left[ { \pm \left( {0.000 \pm 2.416 } \right) + i \cdot \left( {6.572 \pm 0.501} \right)} \right]{\text{fm}}$$ and $$r_{0,\eta ^3He }  = \left[ {  \left( {0.000 \pm 2.416 } \right) + i \cdot \left( {1.268 \pm 0.212} \right)} \right]{\text{fm}}\,.
$$
The scattering length and the effective range are thus determined by
the imaginary parts alone. The fit results don't
fulfil the criterion $|a_r|>|a_i|$.
The ANKE collaboration \cite{Mersmann07} applied in addition another
fitting form. They assumed a two pole representation of the final
state interaction
\begin{equation}
f_s(p)=\frac{f_B}{(1-\frac{p}{p_1})(1-\frac{p}{p_2})}\,
\label{equ:poles}
\end{equation}
with $p_1$ and $p_2$ two complex pole positions. From the position of the first pole one gets scattering length and effective range.
which are $$
a_{\eta ^3He }  = \left[ { \pm \left( {10.7 \pm 0.8_{ - 0.5}^{ + 0.1} } \right) + i \cdot \left( {1.5 \pm 2.6_{ - 0.9}^{ + 1.0} } \right)} \right]{\text{fm}}$$ and $$
r_{0,\eta ^3He }  = \left[ {\left( {1.9 \pm 0.1} \right) + i \cdot \left( {2.1 \pm 0.2_{ - 0.0}^{ + 0.2} } \right)} \right]{\text{fm}}\,.$$
In obtaining these values a smearing of the energy scale due to a
finite beam momentum distribution was applied. This results in a
pole (if exists) at $B_\eta=0.30\pm 0.15 \pm 0.04$ MeV and
$\Gamma_\eta/2 = 0.21 \pm 0.29\pm 0.6$ MeV.

Although
the \fsi parameters differ drastically from those of the fit
the two fit curves are practically
indistinguishable especially in the strong rising  part which is
decisive for the scattering length. It is somewhat surprising that
two fits with five parameters each and a one to one correspondence
give so different results.

Measurements of reaction $d+d\to \eta+^4$He were reported in \cite{Frascaria94}, \cite{Willis97}, \cite{Wronska05} and more recently in \cite{Budzanowski09b}. The cross section is much smaller than for the previously discussed reaction $d+p\to \eta+^3$He.

In a simultaneous analysis of the $d+p\to ^3$He$+\eta$ reaction and the $d+d\to ^4$He$+\eta$ reaction in terms of a simple optical model approach \cite{Willis97} it was found that for the possible binding energies the relation $B_\eta(^3\text{He}+\eta)<B_\eta(^4\text{He}+\eta)$ holds. However, in the measurements with a polarised beam on which the analysis was based, the full polar angle could not be measured. The $s$-wave cross section was extracted by assuming isotropic emission. This isn't true for the data measured at the higher momenta as can be seen by comparing to the data from Ref. \cite{Wronska05}. The anisotropy could be either due to $s$ waves plus $p$ waves or to a $s-d$ wave interference. The problem could be only solved by applying polarised deuterons. Such an experiment was performed by the GEM collaboration at COSY J\"{u}lich \cite{Budzanowski09b} which will be discussed now in some detail.

The experiment was performed at a deuteron beam momentum of 2385.5 MeV/c corresponding to an excess energy of 16.6 MeV \cite{Budzanowski09b}. Recoiling $\alpha$ particles were identified and their four momentum vector measured with the magnetic spectrograph Big Karl \cite{Drochner98}. The experiment made use of polarised as well as unpolarised deuteron beams. The experiment had certain acceptances so that the polarised cross section depends practically only on the analysing power $A_{xx}$.
\begin{figure}[ht]
\begin{center}
\includegraphics[width=0.5\textwidth]{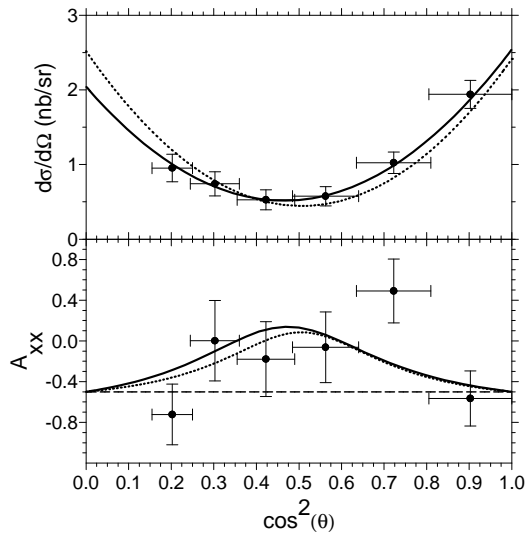}
\caption{Angular distributions of the unpolarised cross section and the analysing power $A_{xx}$ (from Ref. \cite{Budzanowski09b}). The solid
curves represent a fit with four partial waves; the dotted curves show fits with helicity amplitudes.}
\label{Fig:sig-A_xx}
\end{center}
\end{figure}
The angular distribution of the tensor analysing power and of the unpolarized cross section are shown in Fig. \ref{Fig:sig-A_xx}. The $s$ wave amplitude $f_s$ can now be extracted.
We find that $|f_s|^2 = 4.4\pm 1.1$ nb/sr.

For the two highest energy measurements from Ref. \cite{Willis97} $d$ wave contributions have to be considered. With the present result previous analysis could be corrected lading   It is a good approximation to assume the $d$ wave amplitudes $A_2$ and $B$ to depend on the $\eta$ momentum as $p_\eta^2$ and apply the results discussed here. This yields $|f_s|^2 = 13.8\pm 1.2$ nb/sr and $|f_s|^2 = 10.6\pm 1.3$ nb/sr for the momenta at 73 MeV/c and 91 MeV/c, respectively. For the Wronska result we find $|f_s|^2 = 14.3\pm 2.4$ nb/sr at 86 MeV/c.
We are now in a position to make a comparison of the world data for the $s$ wave amplitude. This is done in Fig. \ref{Fig:Compare_he3-He4}. In a fit the production amplitude and the scattering length were fitted to the data yielding $a_{\eta\alpha}=[\pm (3.1\pm 0.5)+i(0.0\pm 0.5)]$ fm. This result corresponds to a bound state - if it exists - of $B_\eta=3.71\pm 0.09$ MeV and $\Gamma/2=0.0\pm 0.2$ MeV. In this Figure we compare the excitation functions for the present reaction with the one for $d+p\to \eta+^3$He. The latter reaction shows a much more rapid rise than the former.
This is an indication of the larger scattering length in case of the lighter system.

Two experiments have been reported leading to the mirror nucleus $^7$Be. The experiments were performed at SATURNE
Saclay \cite{Scomparin93} and COSY J\"{u}lich \cite{Budzanowski10}. Both studies employed the reaction
\begin{equation}\label{equ:Be7}
p+{^6\text{Li}}\to \eta+{^7\text{Be}}\,.
\end{equation}
At Saclay the $\eta$ was measured through its two $\gamma$ decay at a beam energy of 683 MeV corresponding to a beam momentum of 1322 MeV/c or to an excess energy of $Q=19.13$ MeV.
In total eight events were observed. Four excited states with $L=1$ and $L=3$ can contribute. The other experiment \cite{Budzanowski10} was performed at a beam energy of 673.1 MeV, corresponding to 1310 MeV/c momentum or an excess energy of $Q=11.28$ MeV. The recoiling $^7$Be nuclei were detected in the spectrograph Big Karl. Since the $L=3$ states are particle unstable only the two $L=1$ states contribute. The standard detectors in the focal plane were not adequate for this experiment since the recoiling particles have rather low energies of $\approx$100 MeV. The MWDC's were replaced by multi-wire avalanche-chambers to measure the track, followed by two layers of scintillation detectors one metre apart. They allow particle identification via TOF measurement. All these devices were housed in a large vacuum box made of stainless steel.
\begin{figure}[h]
\begin{center}
\includegraphics[width=0.5\textwidth]{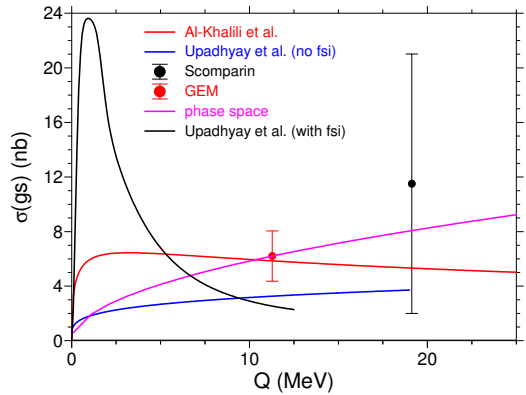}
\caption{Excitation function for the $p+^6\text{Li}\to\eta +
^7\text{Be(g.s.)}$ reaction. The two data points are from Refs.
\cite{Scomparin93} and \cite{Budzanowski10}. The dashed curve is the
Al-Khalili model \cite{Al-Khalili93} and the dashed-dotted is the  phase space behaviour, both
normalised to the GEM data point. The solid and the dotted curves
are calculations \cite{Upadhyay09} with and without final state
interactions. The arrow indicates the region where only the ground state is involved in the reaction.}
\label{Fig:exfu_be7}
\end{center}
\end{figure}

The $\eta$ meson events were identified via the missing mass technique. Finally the counts were converted to cross section. Assuming isotropic emission one gets the total cross section shown in Fig. \ref{Fig:exfu_be7}.
Together with the form factors from Ref. \cite{Al-Khalili93} and the
amplitude $f(pd\to\eta ^3\text{He})$ extracted from the two data sets discussed above, the cross section for the reaction leading to the $^7$Be ground state could be extracted. The two data are shown in Fig. \ref{Fig:exfu_be7}. Also shown is the energy dependence of the Al-Khalili model \cite{Al-Khalili93} normalised to the cross section of the GEM collaboration \cite{Budzanowski10}. Also the normalised phase space dependence is shown. In addition model predictions \cite{Upadhyay09} with and without \fsi are shown. A measurement even closer to threshold preferably below the first excited state could distinguish between the different models and could answer whether strong \fsi exists in this final channel. Upadhyay et al. \cite{Upadhyay09} got from $a_{\eta N} =(0.88+i0.41)$ fm  a value $a_{\eta ^7\text{Be}} = (-9.18+i8.53)$ fm.

%
%%\clearpage
%%%%%%%%%%%%%%%%%%%%%%%%%%%%%%%%%%%%%%%%%%%%%%%%%%%%%%%%%%%%%%%%%%%%%
%%%%%%%%%%%%%%%%%%%%%%%%%%%%%%%%%%%%%%%%%%%%%%%%
% Begin References
%%%%%%%%%%%%%%%%%%%%%%%%%%%%%%%%%%%%%%%%%%%%%%%%

\end{document}